\documentclass[conference]{IEEEtran}
\ifCLASSOPTIONcompsoc
\else
\fi

\ifCLASSINFOpdf
\else
\fi

\usepackage{cite}
\usepackage{url}
\usepackage{minted}
\usepackage{wrapfig}
\usepackage[pdftex]{graphicx}
\usepackage{float}
\usepackage{algorithmic}
\usepackage[tight,footnotesize]{subfigure}
\usepackage{booktabs}
\usepackage{caption}
\usepackage{adjustbox}
\usepackage{makecell} 
\usepackage[linesnumbered,ruled]{algorithm2e}
\usepackage[left=1.62cm,right=1.62cm,top=1.9cm]{geometry}
\usepackage{pifont}
\newcommand{\cmark}{\ding{51}}%
\newcommand{\xmark}{\ding{55}}%

\begin{document}



\title{An Agentic AI Control Plane for 6G Network Slice Orchestration, Monitoring, and Trading}

\author{
\IEEEauthorblockN{
Eranga Bandara\IEEEauthorrefmark{1},
Ross Gore\IEEEauthorrefmark{1},
Sachin Shetty\IEEEauthorrefmark{1},
Ravi Mukkamala\IEEEauthorrefmark{1},
Tharaka Hewa\IEEEauthorrefmark{6}
Abdul Rahman\IEEEauthorrefmark{2},\\
Xueping Liang\IEEEauthorrefmark{3},
Safdar H. Bouk\IEEEauthorrefmark{1},
Amin Hass\IEEEauthorrefmark{7},
Peter Foytik\IEEEauthorrefmark{1},
Ng Wee Keong\IEEEauthorrefmark{4},
Kasun De Zoysa\IEEEauthorrefmark{5}
}

\IEEEauthorblockA{
\IEEEauthorrefmark{1}
{\rm\{cmedawer, sbouk, mukka, rgore, sshetty \}@odu.edu}\\
Old Dominion University, Norfolk, VA, USA\\
}
\IEEEauthorblockA{\IEEEauthorrefmark{2}
{\rm abdulrahman@deloitte.com}\\
Deloitte \& Touche LLP \\
}
\IEEEauthorblockA{\IEEEauthorrefmark{3}
{\rm xuliang@fiu.edu}\\
Florida International University, USA\\
}
\IEEEauthorblockA{\IEEEauthorrefmark{6}
{\rm tharaka.hewa@oulu.fi}\\
Center for Wireless Communications, University of Oulu, Finland\\
}
\IEEEauthorblockA{\IEEEauthorrefmark{7}
{\rm amin.hassanzadeh@accenture.com}\\
Accenture Technology Labs, Arlington, VA, USA\\
}
\IEEEauthorblockA{\IEEEauthorrefmark{4}
{\rm AWKNG@ntu.edu.sg }\\
Nanyang Technological University, Singapore\\
}
\IEEEauthorblockA{\IEEEauthorrefmark{5}
{\rm\{kasun\}@ucsc.cmb.ac.lk}\\
University of Colombo School of Computing, Sri Lanka\\
}
}

\maketitle

\begin{abstract}

6G networks are expected to be AI-native, intent-driven, and economically programmable, requiring fundamentally new approaches to network slice orchestration. Existing slicing frameworks, largely designed for 5G, rely on static policies and manual workflows and are ill-suited for the dynamic, multi-domain, and service-centric nature of emerging 6G environments. In this paper, we propose an agentic AI control plane architecture for 6G network slice orchestration, monitoring, and trading that treats orchestration as a holistic control function encompassing slice planning, deployment, continuous monitoring, and economically informed decision-making. The proposed control plane is realized as a layered architecture in which multiple cooperating AI agents. 
To support flexible and on-demand slice utilization, the control plane incorporates market-aware orchestration capabilities, allowing slice requirements, pricing, and availability to be jointly considered during orchestration decisions. A natural language interface, implemented using the Model Context Protocol (MCP), enables users and applications to interact with control-plane functions through intent-based queries while enforcing safety and policy constraints. To ensure responsible and explainable autonomy, the control plane integrates fine-tuned large language models organized as a multi-model consortium, governed by a dedicated reasoning model. 
The proposed approach is evaluated using a real-world testbed with multiple mobile core instances (e.g Open5GS) integrated with Ericsson’s RAN infrastructure. The results demonstrate that combining agentic autonomy, closed-loop SLA assurance, market-aware orchestration, and natural language control enables a scalable and adaptive 6G-native control plane for network slice management, highlighting the potential of agentic AI as a foundational mechanism for future 6G networks.

\end{abstract}


\begin{IEEEkeywords}
6G , Agentic AI , Responsible AI , Explainable AI , LLM , Model Context Protocol
\end{IEEEkeywords}

\section{Introduction}

6G networks are expected to be AI-native, intent-driven, and economically programmable, requiring fundamentally new approaches to network slice orchestration. As 6G targets immersive applications, autonomous systems, integrated sensing and communication, and AI-driven services, network slicing must evolve beyond static resource partitioning toward adaptive, service-centric control~\cite{agentic-ai-6g}. These emerging requirements challenge existing slice management solutions, which were largely designed for 5G and rely on predefined templates, manual workflows, and limited feedback loops. A key limitation of current approaches is the fragmented treatment of slice lifecycle functions---orchestration, monitoring, and commercial considerations---are often handled by separate systems, resulting in delayed reactions to changing network conditions and inefficient resource utilization. In addition, the growing heterogeneity of 6G infrastructures spanning cloud, edge, radio access, and core networks makes manual or rule-based orchestration increasingly impractical. Continuous adaptation, policy compliance, and cross-domain coordination are therefore essential, yet difficult to achieve with conventional control mechanisms~\cite{5g-slice-overview, slice-gpt}.

At the same time, recent advances in \emph{agentic AI systems}, in which multiple autonomous agents cooperate to plan, execute, and adapt actions in pursuit of high-level goals~\cite{agentic-ai}. Such systems are well-suited for 6G slice management, as they can reason over intent, constraints, and runtime conditions while supporting closed-loop control at scale. Moreover, 6G introduces an increased emphasis on \emph{economic programmability}, where network slices are treated as on-demand digital services whose allocation is influenced not only by technical constraints but also by pricing, availability, and service-level objectives~\cite{ai-native-6g}. These trends motivate the need for an \emph{AI-native control plane} that integrates technical decision-making, continuous monitoring, and market-aware resource allocation within a unified operational entity.

Motivated by these challenges, this paper proposes an \textbf{agentic AI control plane for 6G network slice orchestration, monitoring, and trading}. The proposed control plane treats orchestration as a holistic control function that includes slice planning and deployment, continuous monitoring of SLA, and economically informed decision-making. It is realized as a layered architecture in which cooperating AI agents autonomously interpret user intents, generate slice deployment plans, execute slice lifecycle operations on Kubernetes-based infrastructures, and monitor SLA compliance using real-time telemetry~\cite{6g-slo, agentic-ai}. A natural language interface, implemented using the Model Context Protocol (MCP)~\cite{mcp, mcc}, enables safe and expressive intent-based interaction with control-plane functions while enforcing policy and safety constraints. To ensure responsible and explainable autonomy, the control plane integrates fine-tuned large language models organized as a multi-model consortium with a dedicated reasoning layer~\cite{reasoning-llms} that consolidates model outputs, enforces governance constraints, and provides traceable explanations for orchestration actions~\cite{llama-recipe}. The proposed approach is evaluated using a real-world testbed with multiple mobile core instances (Open5GS) integrated with Ericsson’s next-generation RAN infrastructure~\cite{open5gs-srs}. By unifying agentic autonomy, closed-loop assurance, market awareness, and explainable AI within a single control plane, this work presents a 6G-native approach to network slice management and demonstrates how agentic AI can serve as a foundational control mechanism for scalable, adaptive, and economically efficient network slicing in future 6G networks. The contributions of this paper are as follows:

\begin{enumerate}
    \item We propose a novel agentic AI control plane for 6G network slice orchestration, monitoring, and trading, treating orchestration as a holistic control function that integrates slice planning, deployment, continuous SLA monitoring, and market-aware decision-making.

    \item We design a layered control-plane architecture with cooperating AI agents that autonomously interpret user intents, generate slice deployment plans, execute slice lifecycle operations on Kubernetes-based infrastructures, and adapt to runtime conditions through closed-loop SLA assurance.

    \item We introduce a responsible and explainable AI control-plane design based on fine-tuned large language models organized as a multi-model consortium with a dedicated reasoning layer, enabling governed, auditable, and explainable orchestration decisions in safety-critical 6G environments.

    \item We validate the proposed control plane through a real-world testbed deployment integrating multiple mobile core instances (Open5GS) with next-generation RAN infrastructure, demonstrating the feasibility and effectiveness of agentic AI–driven slice orchestration as a foundation for future 6G networks.
\end{enumerate}

\begin{figure}[t]
\centering{}
\includegraphics[width=3.5in]{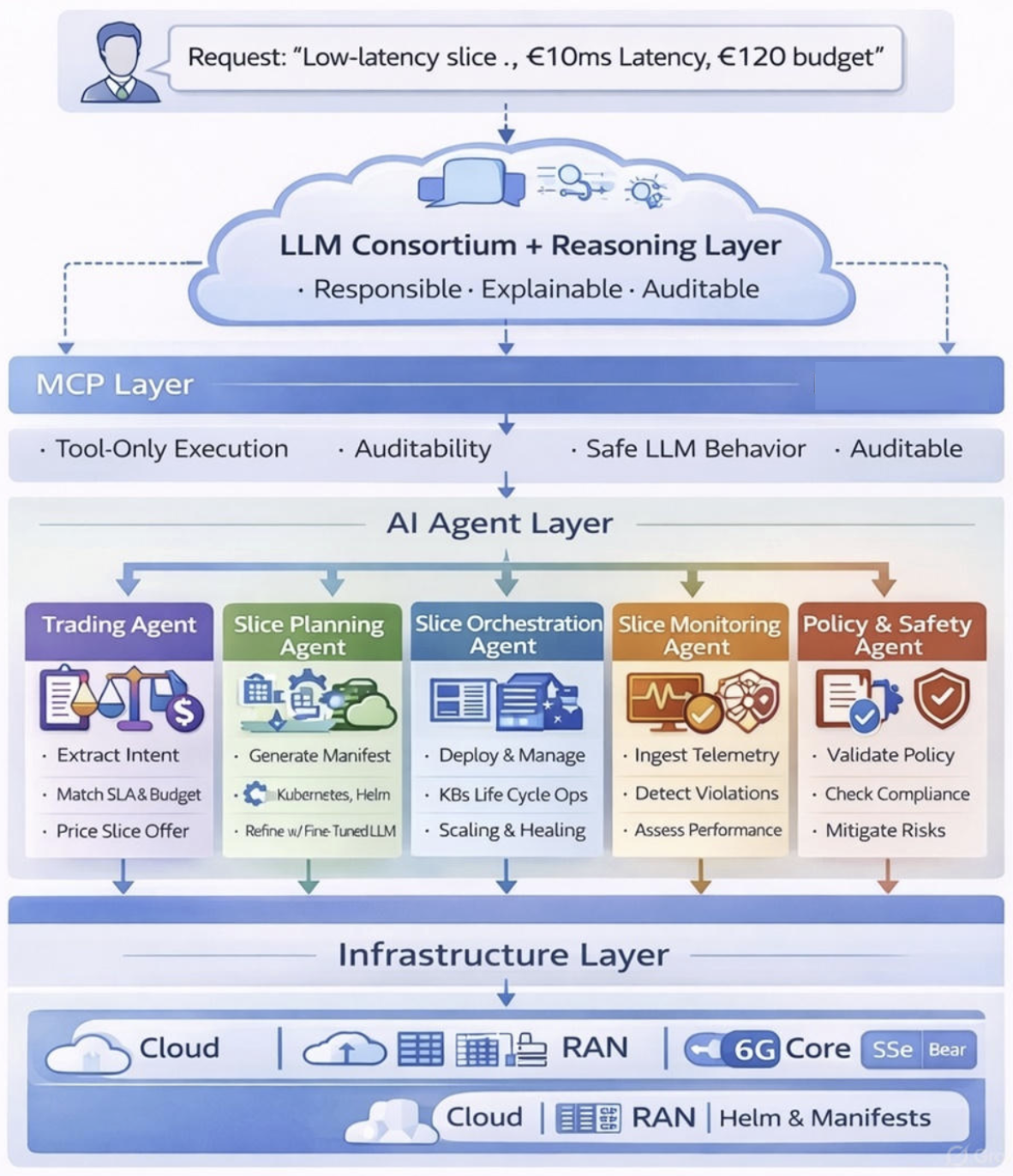}
\DeclareGraphicsExtensions.
\caption{Layered architecture of the proposed agentic AI control plane for 6G network slice orchestration, monitoring, and trading.}
\label{indy528-architecture}
\end{figure}

\section{System Architecture}

Figure~\ref{indy528-architecture} illustrates the architecture of the proposed agentic AI control plane for 6G network slice orchestration, monitoring, and trading. The control plane is organized into five logical layers, each responsible for a distinct set of functions while collectively enabling end-to-end, closed-loop slice control across heterogeneous 6G infrastructures. This layered design promotes modularity, extensibility, and a clear separation of concerns between user interaction, intelligent decision-making, and infrastructure execution. The five layers of the control plane are the \textit{User Layer}, \textit{MCP Layer}, \textit{LLM Layer}, \textit{AI Agents Layer}, and \textit{Infrastructure Layer}. The responsibilities and interactions of each layer are described below.

\subsection{Infrastructure Layer}

The Infrastructure Layer represents the underlying 6G networking and computing resources on which network slices are instantiated and operated. It comprises heterogeneous resources spanning cloud and edge computing platforms, radio access networks (RAN), and 6G core network functions~\cite{srs-ran, bassa-llama}. These resources may be distributed across multiple administrative domains and geographical regions. In the proposed framework, network slices are realized as cloud-native services deployed on Kubernetes clusters, enabling flexible lifecycle management, scalability, and portability. The infrastructure layer exposes abstracted resource and capability information---such as compute capacity, networking features, and regional availability---to the upper layers, while hiding low-level implementation details. This abstraction enables higher-level orchestration decisions without requiring direct interaction with hardware-specific configurations~\cite{llama-recipe}.

\subsection{AI Agents Layer}

The AI Agents Layer constitutes the autonomous control plane of the proposed framework and is composed of five cooperating agents, each responsible for a specific aspect of network slice orchestration and management. Together, these agents implement goal-driven, closed-loop control over the slice lifecycle~\cite{agentic-workflow-practicle-guide}. The Trading Agent interprets user intents, determines slice configurations, and reasons about pricing and availability, while the Policy and Safety Agent enforces governance, security, and operational constraints~\cite{5g-attack-deep-learning}. The Slice Orchestration Planning Agent translates approved slice requests into deployable blueprints, which are executed by the Slice Orchestration Agent through slice deployment and lifecycle operations on Kubernetes clusters. Finally, the SLA Monitoring Agent continuously evaluates slice performance and SLA compliance using runtime telemetry. The agents communicate through well-defined interfaces and shared state, enabling coordinated decision-making while preserving functional independence~\cite{6g-sla}.

\subsection{LLM Layer}

The LLM Layer provides the language understanding and reasoning capabilities required by the AI agents, and is explicitly designed to support Responsible AI (RAI) and Explainable AI (XAI)~\cite{responsible-ai, xai}. In this work, LLMs are used for two high-impact orchestration tasks: (i) generating Kubernetes deployment manifests (e.g., Helm values and resource specifications) from approved slice requirements, and (ii) mapping natural language user requests into structured MCP tool/function invocations. To improve determinism and reduce hallucinations in these operational tasks, the framework employs \textit{fine-tuned LLMs} specialized for manifest generation and tool-call mapping.

To further strengthen robustness, explainability, and governance, the LLM layer is implemented as a \textit{multi-model consortium} in which multiple heterogeneous LLMs independently generate candidate outputs from the same canonical prompt and shared context~\cite{llama-recipe}. Intermediate candidate outputs are preserved as first-class artifacts, exposing agreement and disagreement across models as an inherent source of explainability. A dedicated \textit{reasoning-layer governance LLM (e.g., OpenAI-gpt-oss)}~\cite{reasoning-llms, gpt-oss} then performs structured consolidation over the consortium outputs, resolving conflicts, filtering unsafe or speculative content, enforcing constraints, and producing a final auditable result. This explicit separation between decision generation (consortium) and decision governance (reasoning layer) ensures that orchestration-relevant outputs are explainable by design and responsible by construction.  

\subsection{MCP Layer}

The MCP Layer acts as a secure and structured interaction bridge between the LLM layer, AI agents, and operational system functions. It exposes slice orchestration, monitoring, and trading capabilities as explicit, well-defined tools that can be invoked by the LLM in a controlled manner~\cite{mcp}. In the proposed framework, MCP serves two critical roles. First, it constrains LLM behavior by requiring that all actions be expressed as typed tool calls rather than free-form instructions. Second, it enables auditability and governance by validating tool invocations, enforcing parameter constraints, and recording the complete sequence of actions and intermediate artifacts used to reach a final decision. This design is particularly important when LLM outputs trigger downstream operational changes, as it provides a verifiable control boundary between language-based reasoning and infrastructure actuation. 

\subsection{User Layer}

The User Layer represents human users and external applications that interact with the framework through intent-based interfaces. Users may include network operators, service providers, enterprises, or vertical applications that request new slices, query the SLA status, or initiate market-driven slice acquisition. Interaction occurs primarily through natural language requests that are translated into structured intents and tool calls via the LLM and MCP layers. By abstracting infrastructure complexity behind the agentic control plane, the user layer enables non-expert stakeholders to access advanced 6G slicing capabilities while ensuring that all actions remain policy-compliant, governed, and auditable~\cite{ai-native-6g}.

\section{Platform Functionality}

\begin{figure}[t]
\centering
\includegraphics[width=3.5in]{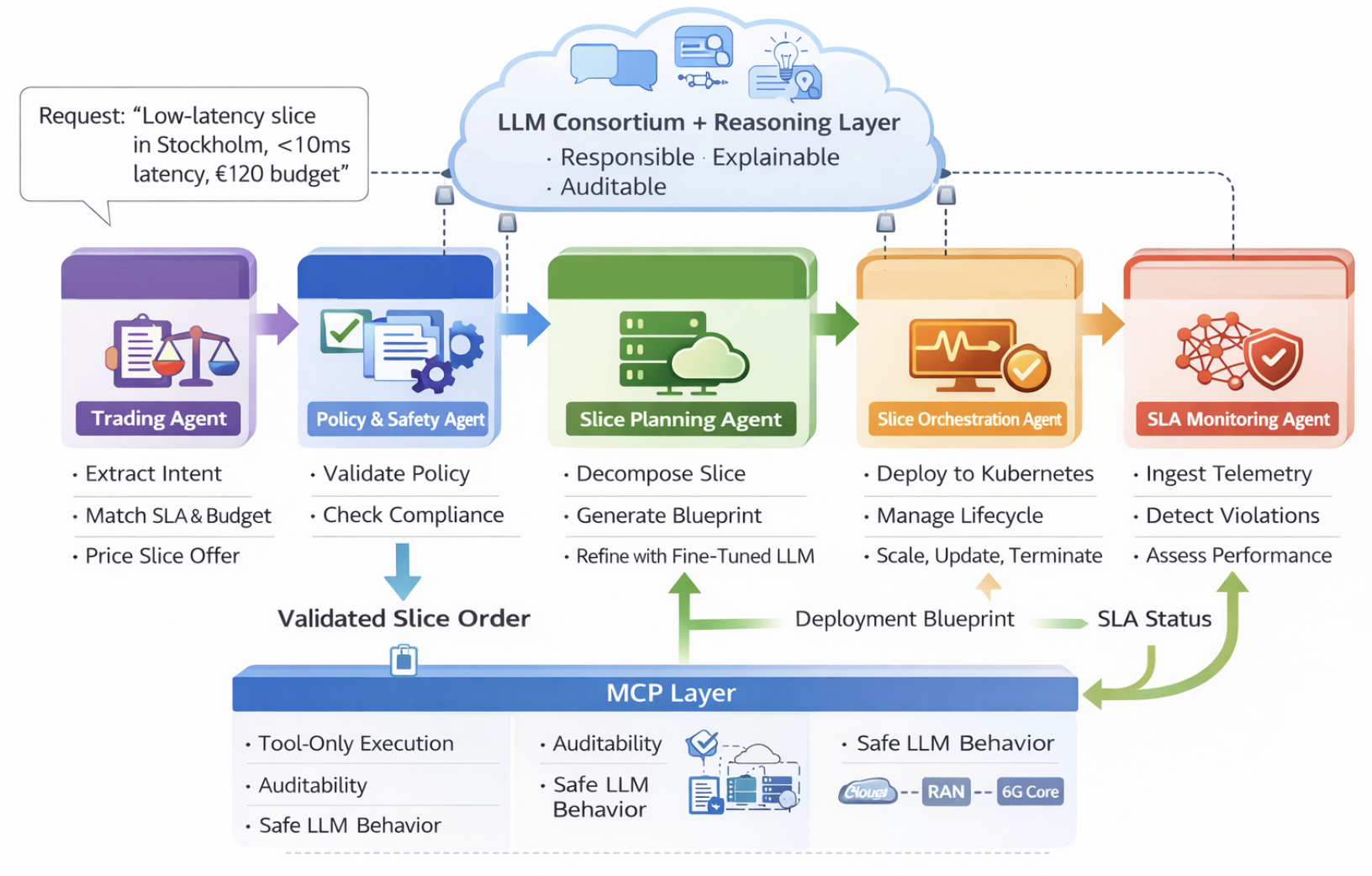}
\DeclareGraphicsExtensions.
\caption{Agentic control-plane flow illustrating intent-based slice orchestration, planning, execution, monitoring, and governance across cooperating AI agents.}
\label{llama2-flow}
\end{figure}

As shown in Figure~\ref{llama2-flow}, the platform is designed around five core functionalities that collectively enable end-to-end, agentic AI--driven 6G network slice orchestration, monitoring, and trading. Each functionality is governed by one or more dedicated AI agents, together forming a closed-loop control plane that manages the complete slice lifecycle from intent specification and economic reasoning to deployment, monitoring, and governance.

\subsection{Intent and Trading Functionality}

The platform supports intent-based slice requests through a natural language interface, enabling users and applications to specify high-level service requirements such as latency targets, geographic scope, duration, and budget constraints without requiring detailed network expertise. These requests are primarily handled by the \emph{Trading Agent}, which interprets the user intent, extracts structured slice requirements, and evaluates them against available slice offerings and current resource availability~\cite{6g-gola-ai}. For example, a user may issue the request: ``\texttt{Provision a low-latency network slice in Stockholm for autonomous vehicle testing for the next two hours, with latency below 10 ms and a maximum cost of €120}.''

The Trading Agent, assisted by the LLM layer, converts this request into a structured intent and invokes MCP tools to retrieve candidate slice offers, pricing information, and feasibility signals. In parallel, the \emph{Policy and Safety Agent} enforces governance constraints, including tenant-specific budget limits, permitted regions, isolation requirements, and compliance policies~\cite{5g-attack-deep-learning}. All LLM-generated actions are expressed as typed MCP tool invocations and validated against schema and policy constraints prior to execution~\cite{slice-leasing}. The outcome of this stage is a validated slice order that captures both service-level objectives and economic terms, which is forwarded to the planning stage.

\subsection{Planning and Manifest Generation}

Once a slice request is approved, the \emph{Slice Orchestration Planning Agent} translates the validated slice order into a deployable slice blueprint. This agent reasons over service requirements, infrastructure constraints, and deployment dependencies to determine the network functions, configurations, and resources required to realize the slice on Kubernetes-based infrastructures.

Fine-tuned large language models are leveraged by the planning agent to generate structured deployment artifacts, such as Kubernetes manifests and Helm configuration values, directly from the validated slice requirements~\cite{slice-gpt}. These artifacts are validated against predefined schemas and deployment policies to ensure correctness, safety, and determinism. The resulting blueprint encodes resource requirements, deployment ordering, and rollback strategies, providing a reproducible plan for the instantiation of slices.

\subsection{Orchestration and Execution}

The \emph{Slice Orchestration Agent} is responsible for executing the deployment blueprint and managing the slice lifecycle on the underlying infrastructure. Using the generated manifests, this agent instantiates slice-related services on target Kubernetes clusters and performs runtime operations such as scaling, modification, healing, and termination~\cite{llama-recipe}.

To ensure reliable execution, the orchestration agent employs staged deployments, continuous health checks, and automated rollback mechanisms in the presence of failures. By abstracting low-level infrastructure interactions, this functionality enables consistent slice lifecycle management across heterogeneous cloud, RAN, and core network environments~\cite{srs-ran}.

\subsection{Monitoring and Assurance}

Runtime monitoring and SLA assurance are handled by the \emph{SLA Monitoring Agent}, which continuously evaluates the performance of active slices against their declared service-level objectives. Telemetry data, including latency, throughput, availability, and resource utilization, is collected from the infrastructure and analyzed in real time.

The SLA Monitoring Agent detects performance degradation and SLA violations, generating alerts and actionable signals that are fed back into the control plane~\cite{6g-sla, ai-native-6g}. These signals enable closed-loop adaptation by triggering corrective actions such as reconfiguration or scaling, ensuring that slices remain aligned with their intended service objectives throughout their operational lifetime.

\begin{figure}[t]
\centering{}
\includegraphics[width=3.5in]{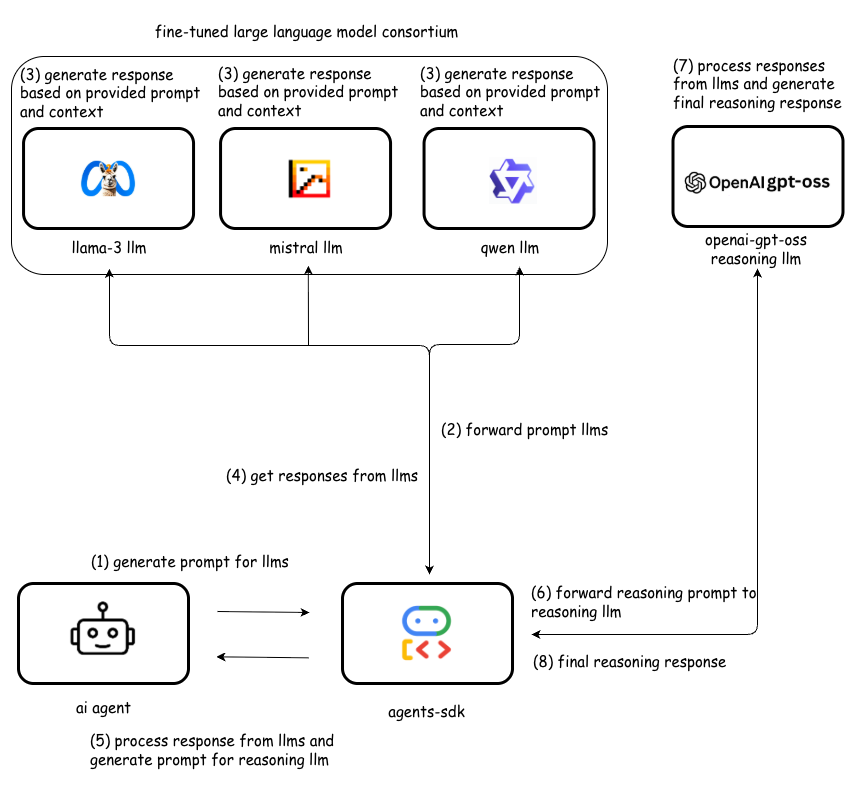}
\DeclareGraphicsExtensions.
\caption{Responsible AI governance architecture based on a multi-model LLM consortium and a reasoning layer.}
\label{responsible-agents-flow}
\end{figure}

\subsection{Responsible AI and Explainable AI Governance}

Given the use of AI-driven automation in safety-critical network operations, the proposed control plane incorporates Responsible and Explainable AI governance as a first-class capability~\cite{towards-rai-xai}. Fine-tuned large language models are deployed as a multi-model consortium, in which multiple heterogeneous models independently generate candidate outputs for planning and control tasks~\cite{llama-recipe, xai}. Rather than relying on a single model decision, this design enables diversity in reasoning and reduces the risk of single-model failure modes. 

For example, during Kubernetes-based slice deployment, three independently fine-tuned LLMs generate candidate deployment manifests from the same validated slice specification, including service class, latency objectives, isolation requirements, and regional constraints. These candidate manifests may differ in resource allocation strategies, placement constraints, or deployment structure. A dedicated reasoning LLM subsequently evaluates the generated artifacts, validates them against orchestration policies and safety constraints, resolves inconsistencies, and selects or synthesizes a final deployment manifest that is both policy-compliant and execution-ready. The interaction between the LLM consortium and the reasoning LLM is illustrated in Figure~\ref{responsible-agents-flow}.

By combining an LLM consortium with centralized reasoning, the proposed governance framework ensures that control-plane decisions remain explainable, auditable, and policy-compliant, while mitigating risks associated with hallucinations, bias, or uncontrolled autonomy and preserving the adaptability and expressiveness of agentic AI~\cite{xai, responsible-ai, agentic-ai}.


\begin{figure}[t]
\centering{}
\includegraphics[width=3.5in]{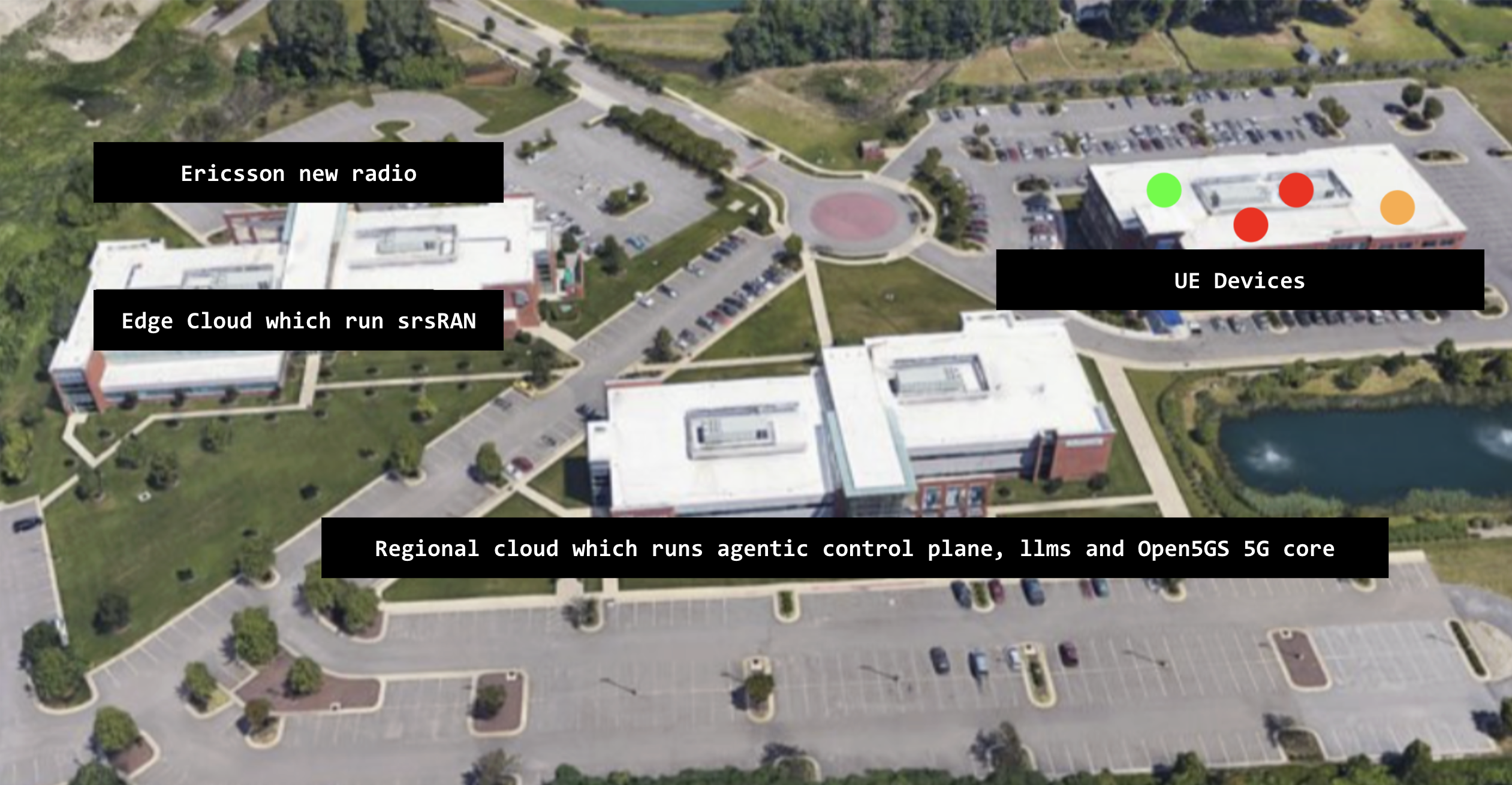}
\DeclareGraphicsExtensions.
\caption{Proposed testbed architecture integrating Ericsson’s next-generation RAN, multiple Open5GS mobile core instances, and on-premise LLM infrastructure deployed at VMASC, Virginia, USA.}
\label{ericsson-tesbed}
\end{figure}

\section{Implementation and Evaluation}

The proposed agentic AI control plane is implemented and evaluated using a real-world sliced network testbed deployed at VMASC, Virginia, USA, integrating multiple Open5GS mobile core network instances with Ericsson’s next-generation RAN infrastructure, as shown in Figure~\ref{ericsson-tesbed}. Network slices are instantiated as cloud-native services on Kubernetes clusters, enabling isolation, scalability, and comprehensive lifecycle management consistent with the proposed architecture~\cite{llama-recipe}. The control plane is realized using the OpenAI Agents SDK~\cite{openai-agent-sdk}, which supports the deployment of modular and cooperative AI agents that autonomously coordinate slice orchestration tasks.

Fine-tuned large language models are employed for two operationally critical functions within the control plane: \emph{(i)} manifest generation and \emph{(ii)} natural language–to–MCP function mapping. Fine-tuning is performed on a AWS \texttt{g5.xlarge} instance equipped with a single GPU (24~GB GPU memory), 4~vCPUs, and 16~GB system memory, using the Unsloth library with LoRA-based parameter-efficient fine-tuning~\cite{llamafactory-unsloth}. Training datasets are structured in JSONL format and consist of paired examples of validated slice specifications and corresponding Kubernetes manifests, as well as natural language intents mapped to schema-compliant MCP tool invocations. To ensure Responsible and Explainable AI behavior, the fine-tuned models are deployed as part of a multi-model LLM consortium, in which multiple heterogeneous models independently generate candidate outputs~\cite{responsible-ai, xai}. A dedicated reasoning layer consolidates these outputs, enforces policy and safety constraints, and produces traceable explanations for accepted orchestration decisions.


The evaluation of the proposed agentic control plane focuses on two complementary aspects: \emph{(i)} the effectiveness of LLM fine-tuning and \emph{(ii)} the predictive performance of the fine-tuned LLMs on the two operationally critical tasks of Kubernetes deployment manifest generation and natural language–to–MCP function mapping. These tasks are selected because they directly influence the correctness, safety, and executability of agentic orchestration actions. Overall, the results demonstrate the feasibility of deploying an agentic AI–driven control plane on contemporary mobile core infrastructure, providing a practical foundation for future 6G network slice management~\cite{slice-gpt}.

\subsection{Evaluation of LLM Fine-Tuning}

\begin{figure}[t]
\centering
\includegraphics[width=3.5in]{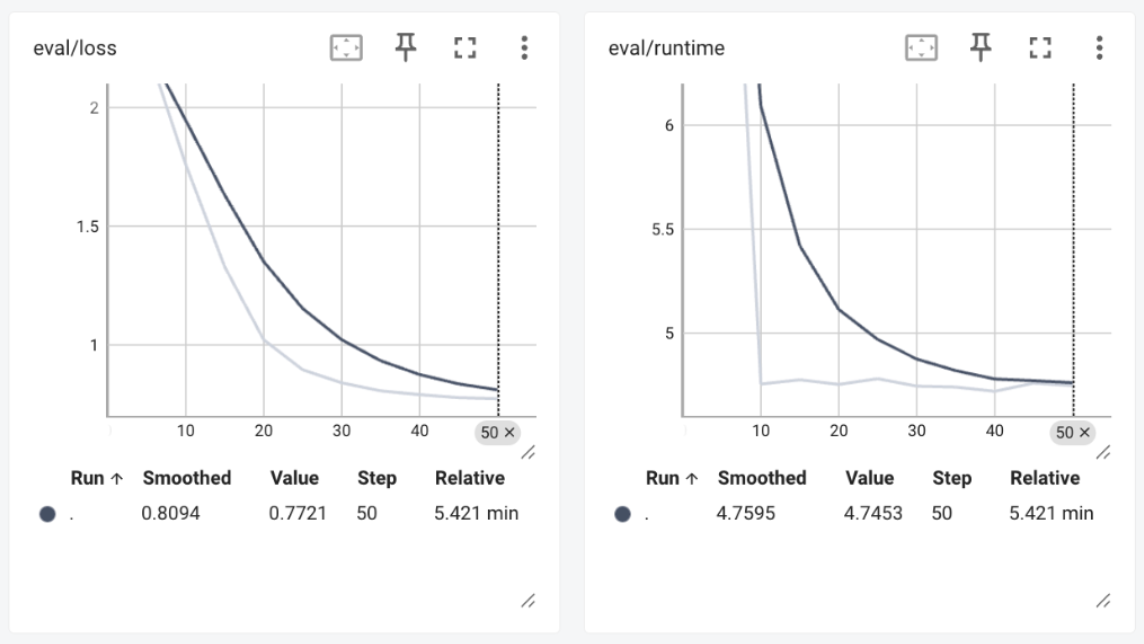}
\DeclareGraphicsExtensions.
\caption{Fine-tuning evaluation metrics showing validation loss convergence and evaluation runtime stability for the Qwen2.}
\label{unsloth-finetune-loss-runtime}
\end{figure}

This evaluation examines the effectiveness of large language model fine-tuning for the proposed control-plane tasks. The fine-tuning results of the Qwen2 LLM~\cite{qwen2} demonstrate stable convergence and effective parameter-efficient adaptation to the target orchestration workloads. As shown in Figure~\ref{unsloth-finetune-loss-runtime}, the validation loss decreases steadily throughout training, converging to a low value by the final step, which indicates improved generalization and the absence of overfitting. In addition, the evaluation runtime remains stable across checkpoints, suggesting predictable inference behavior and suitability for operational deployment within the proposed agentic AI control plane.

\begin{figure}[t]
\centering
\includegraphics[width=3.5in]{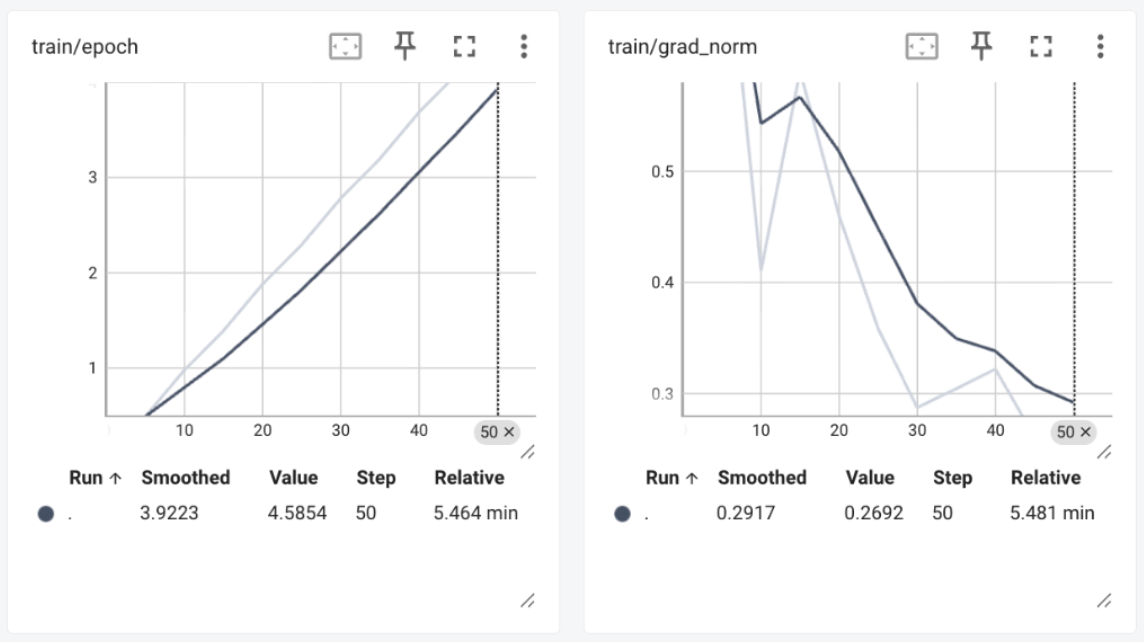}
\DeclareGraphicsExtensions.
\caption{Training dynamics during Qwen2 fine-tuning, illustrating training epoch progression and gradient norm behavior under LoRA-based optimization.}
\label{unsloth-finetune-epoch-grad}
\end{figure}

Complementary training dynamics are illustrated in Figure~\ref{unsloth-finetune-epoch-grad}. The smooth progression of training epochs confirms uninterrupted and efficient optimization, while the gradient norm exhibits a controlled and monotonic decline, indicating well-behaved parameter updates during LoRA-based fine-tuning~\cite{llm-finetune}. Collectively, these results validate the reliability of the fine-tuned Qwen2 model for safety-critical tasks such as Kubernetes deployment manifest generation and natural language–to–MCP function mapping within the proposed agentic AI control plane.

\begin{figure}[t]
\centering
\includegraphics[width=3.2in]{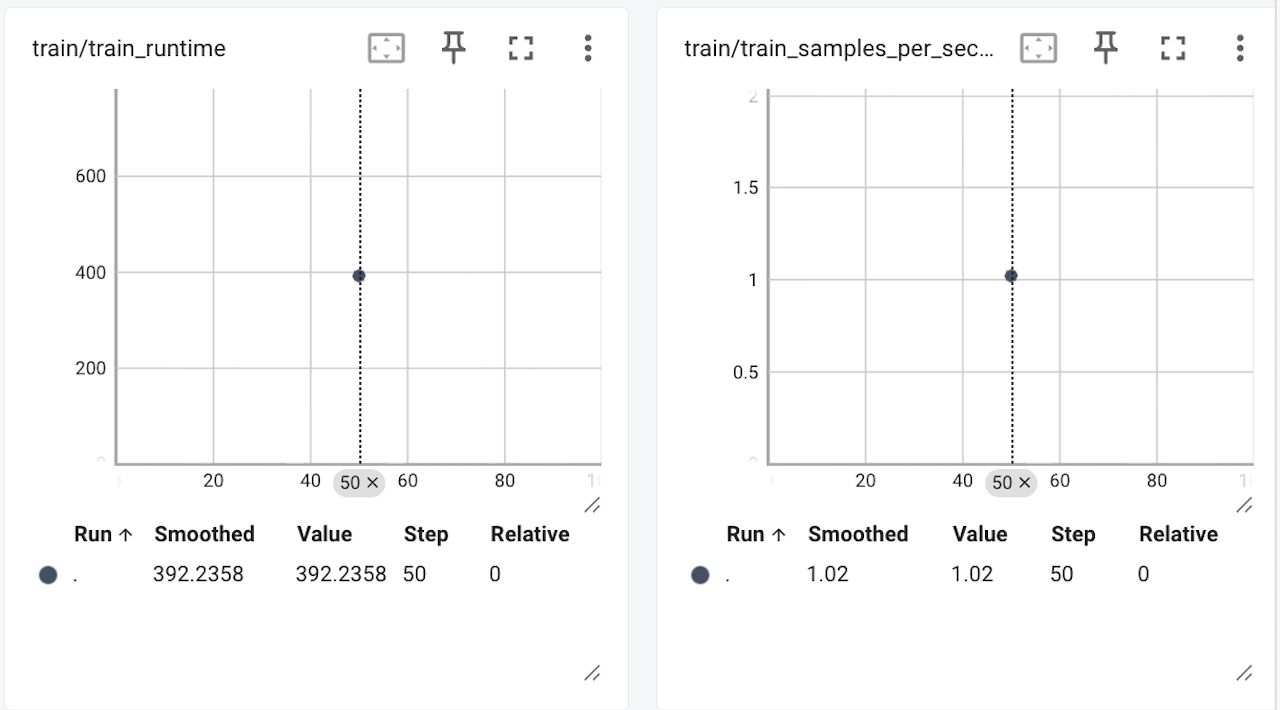}
\DeclareGraphicsExtensions.
\caption{Training efficiency metrics during Qwen2 fine-tuning, showing total training runtime and training throughput (samples per second), illustrating stable and computationally efficient LoRA-based optimization.}
\label{unsloth-finetune-runtime-throughput}
\end{figure}

In addition to convergence behavior, training efficiency metrics further validate the practicality of the fine-tuning process. As shown in Figure~\ref{unsloth-finetune-runtime-throughput}, the total training runtime remains bounded, converging to approximately 392 seconds by the final step, reflecting the computational efficiency of the LoRA-based fine-tuning configuration. The training throughput stabilizes at roughly 1.02 samples per second, indicating consistent resource utilization and a well-balanced data pipeline. These results confirm that the fine-tuning process is not only stable but also operationally efficient, supporting rapid model iteration and deployment within the proposed agentic AI control plane.

\subsection{Predictive Performance of Fine-Tuned LLMs}

\begin{figure}[t]
\centering
\includegraphics[width=3.5in]{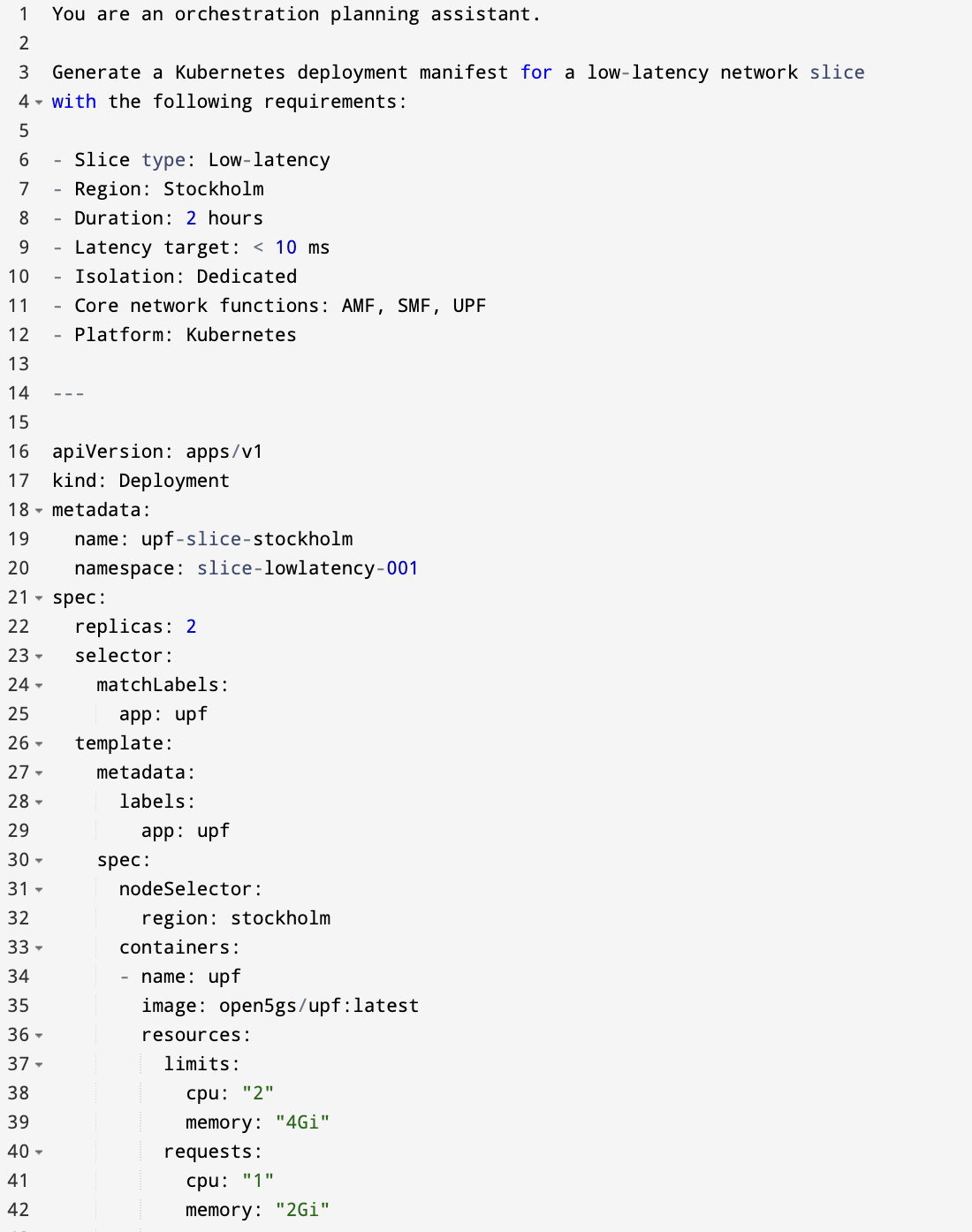}
\caption{Example prompt and corresponding Kubernetes deployment manifest generated by the fine-tuned LLM.}
\label{kube-manifest}
\end{figure}

This evaluation focuses on the predictive performance of the fine-tuned LLMs on two operationally critical tasks: Kubernetes deployment manifest generation and natural language–to–MCP function mapping. For Kubernetes manifest generation, the fine-tuned model is evaluated on its ability to translate validated slice specifications into deployment-ready Kubernetes artifacts that are syntactically correct, semantically aligned with slice objectives, and compliant with orchestration constraints. Figure~\ref{kube-manifest} illustrates an example prompt provided to the fine-tuned model together with the generated Kubernetes manifest. Given slice requirements such as service class, geographic region, latency targets, and isolation levels, the model produces manifests that correctly encode namespace scoping, resource limits, and placement constraints. The generated artifacts can be directly applied to the Kubernetes cluster without manual modification, demonstrating improved determinism and reduced hallucinations compared to non–fine-tuned baselines.

\begin{figure}[t]
\centering
\includegraphics[width=3.5in]{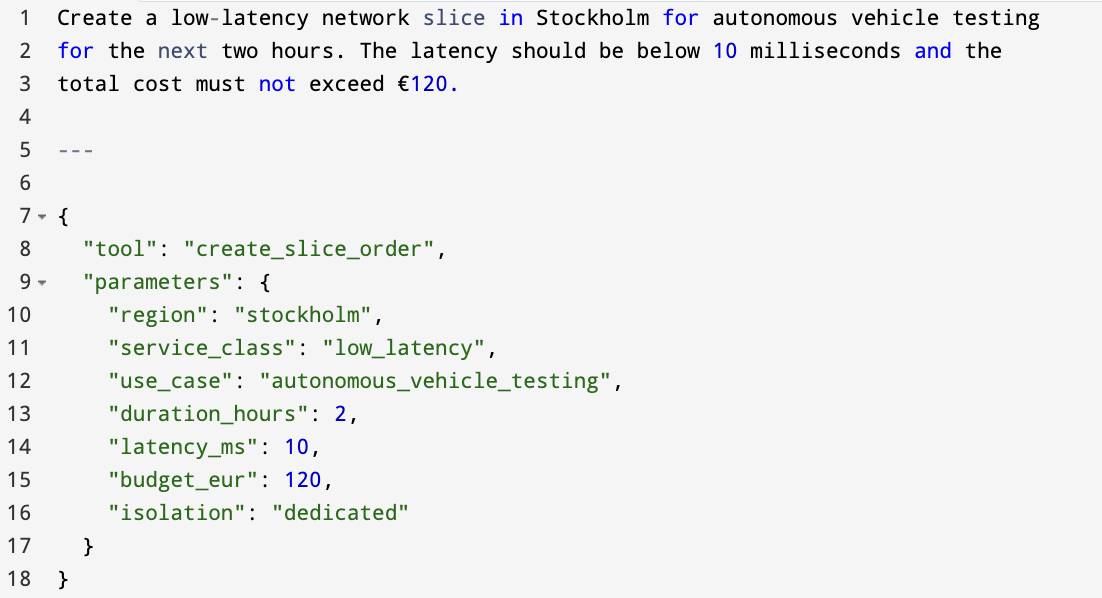}
\caption{Natural language slice request and corresponding MCP function mapping generated by the fine-tuned LLM.}
\label{mcp-mapping-1}
\end{figure}

\begin{figure}[t]
\centering
\includegraphics[width=3.5in]{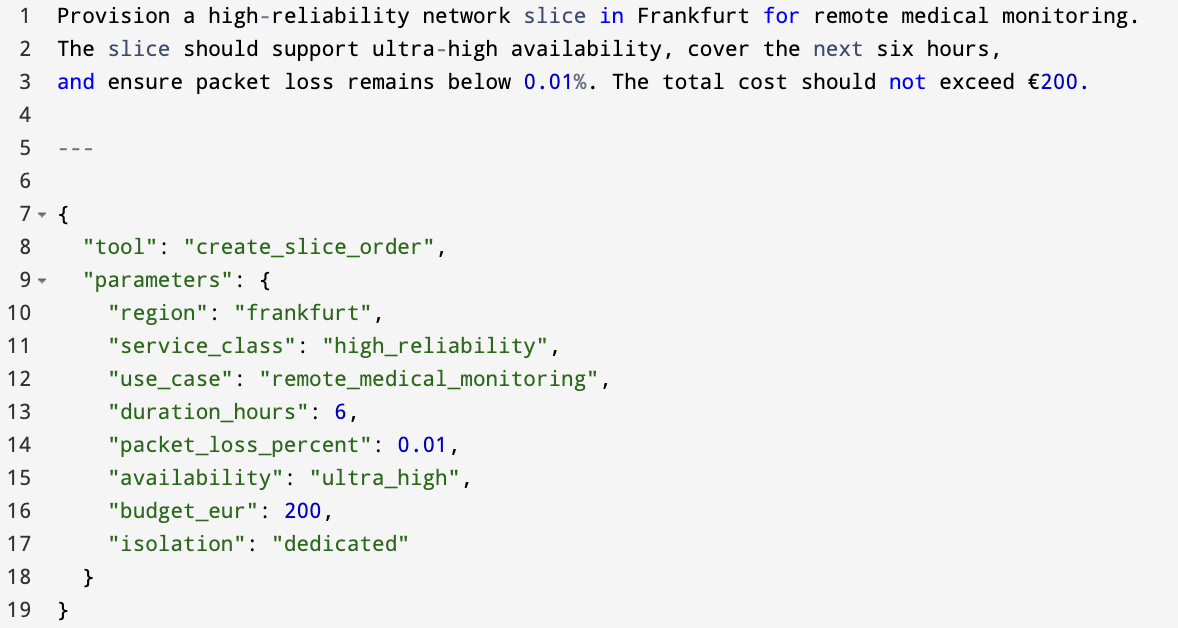}
\caption{Natural language slice request for a high-reliability medical use case and the corresponding MCP function mapping generated by the fine-tuned LLM.}
\label{mcp-mapping-2}
\end{figure}

The next evaluation examines the model’s ability to translate natural language user intents into structured and schema-compliant MCP tool invocations, a capability that is essential for enabling safe and governed natural language control of orchestration functions. Figures~\ref{mcp-mapping-1} and~\ref{mcp-mapping-2} illustrate two representative examples generated by the fine-tuned model, covering latency-sensitive and high-reliability service scenarios. In both cases, the model correctly identifies the appropriate MCP tool and accurately extracts relevant parameters, including region, service class, use case, duration, service-level objectives, and budget constraints. All generated invocations are subsequently validated by the MCP layer prior to execution, ensuring that LLM-driven actions remain policy-compliant, auditable, and safe for deployment within the proposed agentic AI control plane. 

\begin{table*}[!htb]\centering
\caption {Network slice orchestration platform comparison}
\begin{adjustbox}{width=1\textwidth}
\label{t_bc_platforms}
\begin{tabular}{cccccccc}
\toprule

\thead{Platform} & \thead{Centralized/\\Distributed} & \thead{Slice\\Marketplace} & \thead{Responsible-AI\\Support} & \thead{Slice\\ Broker} & \thead{MCP\\Support}  & \thead{Explainable-AI\\ Support} & \thead{LLM fine-tuning\\Support} \\
\midrule
This work & Distributed & \cmark & \cmark & \cmark & \cmark & \cmark & \cmark\\
Blockchain-Slice-Broker~\cite{blockchian-slice-broker} & Distributed & \xmark & \xmark & \cmark & \xmark & \xmark & \xmark\\
\makecell{Network-Slice-Leasing}~\cite{slice-leasing} & Distributed & \xmark & \xmark & \xmark & \xmark & \xmark & \xmark\\
Slice-Market~\cite{network-slice-market} & Distributed & \cmark & \makecell{N/A} & \xmark & \xmark & \xmark & \xmark\\
DBNS & Distributed & \cmark & \xmark & \cmark & \xmark & \xmark & \xmark\\
Network-Slice-Scaling~\cite{network-slice-scaling} & Centralized & \makecell{N/A} & \makecell{N/A} & \xmark & \xmark & \xmark & \xmark\\
Slice-as-a-Service~\cite{slice-as-service} & Centralized & \xmark & \makecell{N/A} & \xmark & \xmark & \xmark & \xmark\\
\bottomrule
\end{tabular}
\end{adjustbox}
\vspace{-0.2in}
\end{table*}

\section{Related Work}

Several studies have investigated end-to-end network slice orchestration with a focus on automation, lifecycle management, and economic models. Blockchain-Slice-Broker~\cite{blockchian-slice-broker} and Network-Slice-Leasing~\cite{slice-leasing} introduce blockchain-enabled slice brokers to support trusted slice trading, while Slice-Market~\cite{network-slice-market} and DBNS~\cite{dbns} extend this concept to decentralized slice marketplaces. Other works, such as Network Slice Scaling~\cite{network-slice-scaling} and Slice-as-a-Service~\cite{slice-as-service}, address slice elasticity and lifecycle management to improve resource utilization and service continuity.

While these approaches advance automation and trust, they largely rely on predefined policies, static templates, or broker-centric control mechanisms. In contrast, this paper proposes an agentic AI–driven control plane that unifies intent-based interaction, autonomous decision-making, continuous SLA monitoring, and market-aware slice management within a single operational entity. The key differences between prior approaches and the proposed control plane are summarized in Table~\ref{t_bc_platforms}, highlighting the shift from static and broker-based designs toward an AI-native, explainable control-plane architecture aligned with future 6G requirements.

\section{Conclusions and Future Work}

This paper presented an agentic AI control plane for 6G network slice orchestration, monitoring, and trading that addresses the emerging requirements of AI-native, intent-driven, and economically programmable 6G networks. By treating orchestration as a holistic control function, the proposed control plane integrates intent-based slice requests, market-aware decision-making, Kubernetes-based slice deployment, continuous SLA monitoring, and natural language interaction within a unified architecture. Its layered design, together with cooperating AI agents, enables scalable and adaptive slice lifecycle management across heterogeneous cloud, RAN, and core network infrastructures. A key contribution of this work is the incorporation of fine-tuned large language models for operationally critical tasks, including Kubernetes deployment manifest generation and natural language–to–MCP function mapping. To ensure trustworthy autonomy, these models are deployed within a multi-model consortium governed by a dedicated reasoning layer that enforces policy constraints and provides explainable and auditable orchestration decisions. Experimental evaluation using a real-world testbed integrating multiple Open5GS mobile core instances with next-generation RAN infrastructure demonstrates that the proposed control plane enables reliable, execution-ready orchestration actions while reducing hallucinations and improving determinism. Overall, this work demonstrates the feasibility and effectiveness of agentic AI–driven control planes as foundational mechanisms for future 6G network slicing. By combining agent autonomy, closed-loop assurance, market awareness, and responsible AI governance, the proposed approach provides a practical pathway toward scalable, adaptive, and economically efficient slice management. Future work will investigate large-scale multi-domain deployments, tighter integration with emerging 6G standards, advanced economic models for dynamic slice marketplaces, and broader evaluation of control-plane robustness under adversarial and highly dynamic network conditions.



\bibliographystyle{IEEEtran}
\bibliography{references}

@article{blockchian-slice-broker,
  title={A blockchain-based network slice broker for 5G services},
  author={Nour, Boubakr and Ksentini, Adlen and Herbaut, Nicolas and Frangoudis, Pantelis A and Moungla, Hassine},
  journal={IEEE Networking Letters},
  volume={1},
  number={3},
  pages={99--102},
  year={2019},
  publisher={IEEE}
}

@article{network-slice-scaling,
  title={Dynamic network slice scaling assisted by prediction in 5G network},
  author={Zhou, Jinhe and Zhao, Wenjun and Chen, Shuo},
  journal={IEEE Access},
  volume={8},
  pages={133700--133712},
  year={2020},
  publisher={IEEE}
}

@article{slice-as-service,
  title={Network slice lifecycle management for 5g mobile networks: An intent-based networking approach},
  author={Abbas, Khizar and Khan, Talha Ahmed and Afaq, Muhammad and Song, Wang-Cheol},
  journal={IEEE Access},
  volume={9},
  pages={80128--80146},
  year={2021},
  publisher={IEEE}
}

@article{5g-slice-overview,
  title={An overview of network slicing for 5G},
  author={Zhang, Shunliang},
  journal={IEEE Wireless Communications},
  volume={26},
  number={3},
  pages={111--117},
  year={2019},
  publisher={IEEE}
}

@inproceedings{slice-leasing,
  title={Blockchain network slice broker in 5G: Slice leasing in factory of the future use case},
  author={Backman, Jere and Yrj{\"o}l{\"a}, Seppo and Valtanen, Kristiina and M{\"a}mmel{\"a}, Olli},
  booktitle={2017 Internet of Things Business Models, Users, and Networks},
  pages={1--8},
  year={2017},
  organization={IEEE}
}

@inproceedings{network-slice-market,
  title={5g network slice brokering: A distributed blockchain-based market},
  author={Afraz, Nima and Ruffini, Marco},
  booktitle={2020 European Conference on Networks and Communications (EuCNC)},
  pages={23--27},
  year={2020},
  organization={IEEE}
}

@article{dbns,
  title={DBNS: A distributed blockchain-enabled network slicing framework for 5G networks},
  author={Togou, Mohammed Amine and Bi, Ting and Dev, Kapal and McDonnell, Kevin and Milenovic, Aleksandar and Tewari, Hitesh and Muntean, Gabriel-Miro},
  journal={IEEE Communications Magazine},
  volume={58},
  number={11},
  pages={90--96},
  year={2020},
  publisher={IEEE}
}

@article{llm,
  title={Language Models Enable Simple Systems for Generating Structured Views of Heterogeneous Data Lakes},
  author={Arora, Simran and Yang, Brandon and Eyuboglu, Sabri and Narayan, Avanika and Hojel, Andrew and Trummer, Immanuel and R{\'e}, Christopher},
  journal={arXiv preprint arXiv:2304.09433},
  year={2023}
}

@article{llamafactory-unsloth,
  title={Llamafactory: Unified efficient fine-tuning of 100+ language models},
  author={Zheng, Yaowei and Zhang, Richong and Zhang, Junhao and Ye, Yanhan and Luo, Zheyan and Feng, Zhangchi and Ma, Yongqiang},
  journal={arXiv preprint arXiv:2403.13372},
  year={2024}
}

@article{qwen2,
  title={Qwen2-vl: Enhancing vision-language model's perception of the world at any resolution},
  author={Wang, Peng and Bai, Shuai and Tan, Sinan and Wang, Shijie and Fan, Zhihao and Bai, Jinze and Chen, Keqin and Liu, Xuejing and Wang, Jialin and Ge, Wenbin and others},
  journal={arXiv preprint arXiv:2409.12191},
  year={2024}
}

@article{5g-attack-deep-learning,
  title={Deep learning-based DDoS-attack detection for cyber--physical system over 5G network},
  author={Hussain, Bilal and Du, Qinghe and Sun, Bo and Han, Zhiqiang},
  journal={IEEE Transactions on Industrial Informatics},
  volume={17},
  number={2},
  pages={860--870},
  year={2020},
  publisher={IEEE}
}

@article{srs-ran,
  title={Multi-UE 5G RAN Measurements: A Gamut of Architectural Options},
  author={Grohmann, Andreas Ingo and Seidel, Mauri and Itting, Sebastian AW and Cheng, Ray-Guang and Reisslein, Martin and Fitzek, Frank HP},
  journal={IEEE Access},
  year={2024},
  publisher={IEEE}
}

@article{reasoning-llms,
  title={Llm as a mastermind: A survey of strategic reasoning with large language models},
  author={Zhang, Yadong and Mao, Shaoguang and Ge, Tao and Wang, Xun and de Wynter, Adrian and Xia, Yan and Wu, Wenshan and Song, Ting and Lan, Man and Wei, Furu},
  journal={arXiv preprint arXiv:2404.01230},
  year={2024}
}

@article{mcp,
  title={A Survey of the Model Context Protocol (MCP): Standardizing Context to Enhance Large Language Models (LLMs)},
  author={Singh, Aditi and Ehtesham, Abul and Kumar, Saket and Khoei, Tala Talaei},
  year={2025},
  publisher={Preprints}
}

@article{agentic-ai,
  title={Agentic AI: Autonomous Intelligence for Complex Goals--A Comprehensive Survey},
  author={Acharya, Deepak Bhaskar and Kuppan, Karthigeyan and Divya, B},
  journal={IEEE Access},
  year={2025},
  publisher={IEEE}
}

@inproceedings{llm-finetune,
  title={Data-efficient Fine-tuning for LLM-based Recommendation},
  author={Lin, Xinyu and Wang, Wenjie and Li, Yongqi and Yang, Shuo and Feng, Fuli and Wei, Yinwei and Chua, Tat-Seng},
  booktitle={Proceedings of the 47th international ACM SIGIR conference on research and development in information retrieval},
  pages={365--374},
  year={2024}
}

@article{open5gs-srs,
  title={Performance evaluation of an open source implementation of a 5G standalone platform},
  author={H{\aa}keg{\aa}rd, Jan Erik and Lundkvist, Henrik and Rauniyar, Ashish and Morris, Peter},
  journal={IEEE Access},
  volume={12},
  pages={25809--25819},
  year={2024},
  publisher={IEEE}
}

@inproceedings{agentic-ai-6g,
  title={The Integration of Agentic AI in 6G Wireless Networks: State-of-the-Art, Challenges, and Future Perspectives},
  author={{\c{C}}{\i}men, Sedat and Karahan, S{\"u}meye Nur and Karhan, Deniz and G{\"u}ll{\"u}, Merve and Osmanca, Mustafa Serdar},
  booktitle={2025 Innovations in Intelligent Systems and Applications Conference (ASYU)},
  pages={1--6},
  year={2025},
  organization={IEEE}
}

@INPROCEEDINGS{slice-gpt,
  author={Bandara, Eranga and Foytik, Peter and Shetty, Sachin and Mukkamala, Ravi and Rahman, Abdul and Liang, Xueping and Keong, Ng Wee and Zoysa, Kasun De},
  booktitle={2024 IEEE 21st Consumer Communications \& Networking Conference (CCNC)}, 
  title={SliceGPT – OpenAI GPT-3.5 LLM, Blockchain and Non-Fungible Token Enabled Intelligent 5G/6G Network Slice Broker and Marketplace}, 
  year={2024},
  volume={},
  number={},
  pages={439-445},
  doi={10.1109/CCNC51664.2024.10454701}}

@INPROCEEDINGS{llama-recipe,
  author={Bandara, Eranga and Bouk, Safdar H. and Shetty, Sachin and Roy, Sandip and Mukkamala, Ravi and Rahman, Abdul and Foytik, Peter and Liang, Xueping and Keong, Ng Wee and De Zoysa, Kasun},
  booktitle={2025 IEEE 22nd Consumer Communications \& Networking Conference (CCNC)}, 
  title={Llama-Recipe — Fine-Tuned Meta's Llama LLM, PBOM and NFT Enabled 5G Network-Slice Orchestration and End-to-End Supply-Chain Verification Platform}, 
  year={2025},
  volume={},
  number={},
  pages={1-6},
  doi={10.1109/CCNC54725.2025.10976116}}

@article{ai-native-6g,
  title={AI-native network slicing for 6G networks},
  author={Wu, Wen and Zhou, Conghao and Li, Mushu and Wu, Huaqing and Zhou, Haibo and Zhang, Ning and Shen, Xuemin Sherman and Zhuang, Weihua},
  journal={IEEE Wireless Communications},
  volume={29},
  number={1},
  pages={96--103},
  year={2022},
  publisher={IEEE}
}

@article{responsible-ai,
  title={LLM potentiality and awareness: a position paper from the perspective of trustworthy and responsible AI modeling},
  author={Sarker, Iqbal H},
  journal={Discover Artificial Intelligence},
  volume={4},
  number={1},
  pages={40},
  year={2024},
  publisher={Springer}
}

@article{xai,
  title={Explainable AI (XAI): Core ideas, techniques, and solutions},
  author={Dwivedi, Rudresh and Dave, Devam and Naik, Het and Singhal, Smiti and Omer, Rana and Patel, Pankesh and Qian, Bin and Wen, Zhenyu and Shah, Tejal and Morgan, Graham and others},
  journal={ACM computing surveys},
  volume={55},
  number={9},
  pages={1--33},
  year={2023},
  publisher={ACM New York, NY}
}

@inproceedings{6g-gola-ai,
  title={Goal-oriented and semantic communication in 6G AI-native networks: The 6G-GOALS approach},
  author={Strinati, Emilio Calvanese and Di Lorenzo, Paolo and Sciancalepore, Vincenzo and Aijaz, Adnan and Kountouris, Marios and G{\"u}nd{\"u}z, Deniz and Popovski, Petar and Sana, Mohamed and Stavrou, Photios A and Soret, Beatriz and others},
  booktitle={2024 Joint European Conference on Networks and Communications \& 6G Summit (EuCNC/6G Summit)},
  pages={1--6},
  year={2024},
  organization={IEEE}
}

@article{openai-agent-sdk,
  title={VTutor: An Open-Source SDK for Generative AI-Powered Animated Pedagogical Agents with Multi-Media Output},
  author={Chen, Eason and Lin, Chenyu and Tang, Xinyi and Xi, Aprille and Wang, Canwen and Lin, Jionghao and Koedinger, Kenneth R},
  journal={arXiv preprint arXiv:2502.04103},
  year={2025}
}

@inproceedings{6g-slo,
  title={DRL-Enabled SLO-Aware Task Scheduling for Large Language Models in 6G Networks},
  author={Mekrache, Abdelkader and Ksentini, Adlen and Verikoukis, Christos},
  booktitle={ICC 2025-IEEE International Conference on Communications},
  pages={813--818},
  year={2025},
  organization={IEEE}
}

@inproceedings{6g-sla,
  title={DRL-Enabled SLO-Aware Task Scheduling for Large Language Models in 6G Networks},
  author={Mekrache, Abdelkader and Ksentini, Adlen and Verikoukis, Christos},
  booktitle={ICC 2025-IEEE International Conference on Communications},
  pages={813--818},
  year={2025},
  organization={IEEE}
}

@article{gpt-oss,
  title={gpt-oss-120b \& gpt-oss-20b Model Card},
  author={Agarwal, Sandhini and Ahmad, Lama and Ai, Jason and Altman, Sam and Applebaum, Andy and Arbus, Edwin and Arora, Rahul K and Bai, Yu and Baker, Bowen and Bao, Haiming and others},
  journal={arXiv preprint arXiv:2508.10925},
  year={2025}
}

@article{towards-rai-xai,
  title={Towards Responsible and Explainable AI Agents with Consensus-Driven Reasoning},
  author={Bandara, Eranga and Hewa, Tharaka and Gore, Ross and Shetty, Sachin and Mukkamala, Ravi and Foytik, Peter and Rahman, Abdul and Bouk, Safdar H and Liang, Xueping and Hass, Amin and others},
  journal={arXiv preprint arXiv:2512.21699},
  year={2025}
}

@article{agentic-workflow-practicle-guide,
  title={A Practical Guide for Designing, Developing, and Deploying Production-Grade Agentic AI Workflows},
  author={Bandara, Eranga and Gore, Ross and Foytik, Peter and Shetty, Sachin and Mukkamala, Ravi and Rahman, Abdul and Liang, Xueping and Bouk, Safdar H and Hass, Amin and Rajapakse, Sachini and others},
  journal={arXiv preprint arXiv:2512.08769},
  year={2025}
}

@article{mcc,
  title={Model Context Contracts-MCP-Enabled Framework to Integrate LLMs With Blockchain Smart Contracts},
  author={Bandara, Eranga and Shetty, Sachin and Mukkamala, Ravi and Gore, Ross and Foytik, Peter and Bouk, Safdar H and Rahman, Abdul and Liang, Xueping and Keong, Ng Wee and De Zoysa, Kasun and others},
  journal={arXiv preprint arXiv:2510.19856},
  year={2025}
}

@INPROCEEDINGS{bassa-llama,
  author={Bandara, Eranga and Bouk, Safdar H. and Shetty, Sachin and Gore, Ross and Kompella, Sastry and Mukkamala, Ravi and Rahman, Abdul and Foytik, Peter and Liang, Xueping and Keong, Ng Wee and De Zoysa, Kasun},
  booktitle={2025 International Wireless Communications and Mobile Computing (IWCMC)}, 
  title={Bassa-Llama — Fine-Tuned Meta’s Llama LLM, Blockchain and NFT Enabled Real-Time Network Attack Detection Platform for Wind Energy Power Plants}, 
  year={2025},
  volume={},
  number={},
  pages={330-336},
  doi={10.1109/IWCMC65282.2025.11059647}}

\end{document}